\documentclass[a4paper]{jpconf}

\usepackage{booktabs}       
\usepackage{amsfonts}       
\usepackage{amsmath}
\usepackage[sort&compress,square,comma,numbers]{natbib}
\usepackage{hyperref}
\usepackage{graphicx}
\usepackage{subcaption}
\usepackage{cleveref}
\usepackage{todonotes}
\usepackage{multirow}
\usepackage{lineno}
\graphicspath{ {./img/} }

\begin{document}
\title{Semi-Equivariant GNN Architectures for Jet Tagging}

\author{Daniel Murnane\textsuperscript{1}, Savannah Thais\textsuperscript{2}, Jason Wong\textsuperscript{3}}

\address{
\textsuperscript{1}Lawrence Berkeley National Laboratory,
\textsuperscript{2}Princeton University,
\textsuperscript{3}University of California Berkeley
}

\ead{dtmurnane@lbl.gov}

\begin{abstract}
    Composing Graph Neural Networks (GNNs) of operations that respect physical symmetries has been suggested to give better model performance with a smaller number of learnable parameters. However, real-world applications, such as in high energy physics have not born this out. We present the novel architecture VecNet that combines both symmetry-respecting and unconstrained operations to study and tune the degree of physics-informed GNNs. We introduce a novel metric, the \textit{ant factor}, to quantify the resource-efficiency of each configuration in the search-space. We find that a generalized architecture such as ours can deliver optimal performance in resource-constrained applications.
\end{abstract}

\section{Introduction}
The use of graph neural networks (GNNs) in high energy physics (HEP) analysis has delivered large increases in accuracy, along with a large number of architectural options. Many of these designs, from large numbers of parameters, to complex graph construction and convolution updates, make GNNs resource-hungry and time-consuming to develop. Constraining operations to obey physically meaningful equivariance has emerged as one possible alternative to large overparameterized models. However, previous work has typically compared one state-of-the-art (SotA) GNN classifier, with another - very different - classifier that is fully equivariant (for example in~\cite{lgn}). The effect of equivariance in these studies is ambiguous when many other differences between models obscure a direct comparison. Furthermore, it is unfair to compare outright accuracy when a motivation for equivariance is to obtain more resource-efficient models. 

In this work, we attempt to resolve these issues by introducing a generic architecture VecNet that abstracts-out as hyperparameters (HPs) many of the design choices of SotA GNN classifiers, in order to perform a thorough exploration of the design space. We focus on the use-case of top quark tagging in jet physics. This architecture also easily allows non-equivariant configurations, fully-equivariant configurations, and a tunable admixture of both - \textit{semi-equivariant} configurations, by handling vectors carefully. We also introduce a metric called an \textit{ant factor}. This metric compares the classification power of a model (i.e. how much weight it can lift) with a model's size. This measure allows us to prioritize a subset of resource-efficient HP configurations. It also uncovers a non-trivial admixture of equivariant and non-equivariant channels for the most powerful model given a set of resource constraints.

\section{Jet Tagging}

\begin{figure}[htb!]
    \centering
    \includegraphics[width=0.9\textwidth]{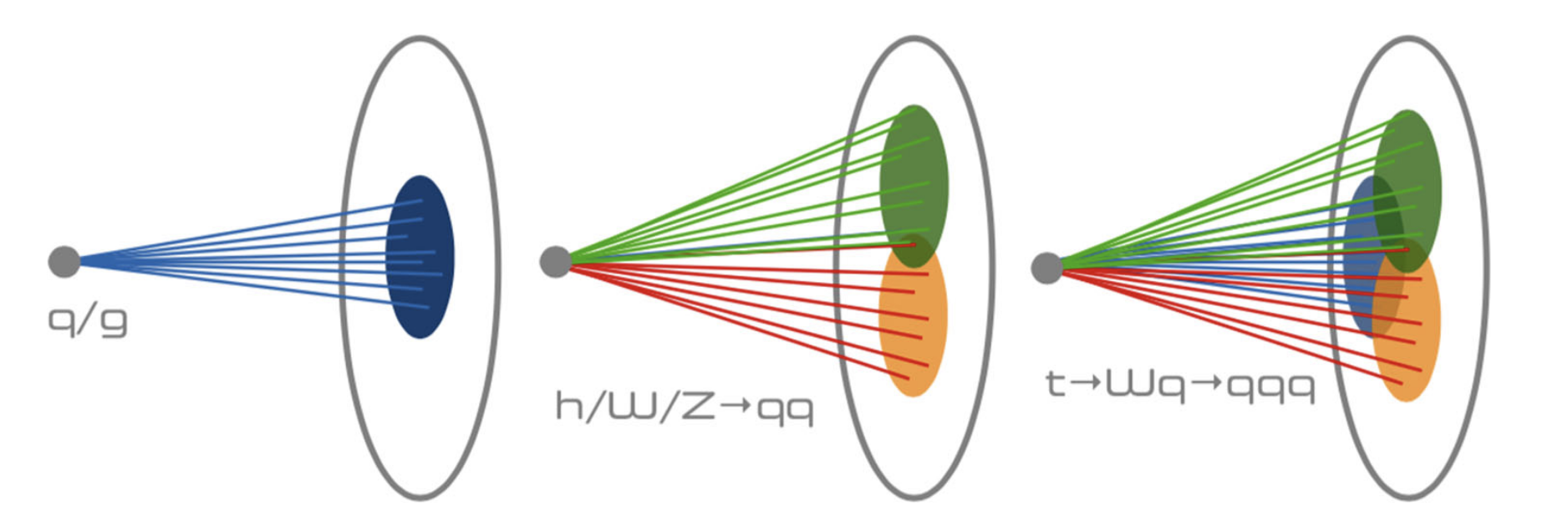} 
    \caption{Pictorial representation of three common types of jets: a single quark/gluon jets (left), a heavy jet decaying to two quarks (center), and a high momentum top quark to heavy boson to quarks decay chain. Figure reproduced from \cite{jedi}}
    \label{fig:jet_diagram}
\end{figure}

Many particles of interest in HEP experiments are produced in association with or decay into quarks and gluons. The fragmentation and hadronization of these quarks and gluons via quantum chromodynamics (QCD) results in collimated sprays of particles called jets. Jet tagging is the problem of identifying what type of particle a jet originated from, and is a critical task at HEP experiments for enabling searches for new particles and precision measurements. This task is complicated by the fact that electroweak particles like vector bosons or heavy theoretical particles that decay into multiple particles are `boosted' due to the high energy of their production, such that their individual hadronic decays are reconstructed as a single jet (see \cref{fig:jet_diagram}). 

Due to the ubiquity, importance, and complexity of jets in HEP experiment data, there has been a broad range of algorithms developed for the task of jet tagging, including multivariate methods based on derived physics quantities~\cite{non-ml_taggers} and more recently machine learning (ML) methods using lower-level information~\cite{ml_btagging,ml_toptagging} (further described in \cref{sec:ml_jets}). There are several important metrics to consider when evaluating these algorithms; physics performance, typically evaluated in terms of accuracy across different jet types and kinematics, is critical. Computing performance, including inference time and resource utilization, is also a key consideration. In particular, jet tagging is an essential component of the trigger system at Large Hadron Collider (LHC) experiments like ATLAS and CMS; the trigger system is responsible for online data-processing that efficiently identifies interesting events to retain for further offline processing and discards the rest. The latency requirements of these systems are already extremely tight and will become even more restrictive after the upgrade to the more data-intensive High Luminosity LHC. For example, the CMS Level 1 Trigger for HL-LHC must process a full event ($\sim$7Mb) in less than 12.5 $\mu$s~\cite{cms_hllhc_trigger}. While hardware-based algorithm acceleration and parallelization on GPUs and FPGAs can help in addressing these computing needs, these requirements also motivate the development of small, highly efficient jet tagging models with higher inference throughput and reduced resource utilization. 

\subsection{Dataset}
In this work we focus specifically on the problem of top quark tagging using the Open Data Top Quark Tagging Reference Dataset \cite{dataset}. This dataset consists of 2M zero pileup events containing either a hadronically decaying top quark (signal) or QCD dijets (background); the dataset is split into 1.2M training events, 40k validation events, and 40k test events. All jets are generated with PYTHIA8 \cite{pythia} and passed through the DELPHES ATLAS detector card for fast detector simulation \cite{delphes}. Jets are clustered using the anti-$k_T$ algorithm \cite{anti-kt} with a distance parameter $R=0.8$ and a required to have a transverse momentum $p_T\in[550,650]$ GeV and a pseudorapidity $|\eta|<2$. Additionally, signal jets are required to be matched to a parton-level top quark within $\Delta R=0.8$ and all three quarks from the top decay must be within $\Delta R=0.8$ of the central jet axis. For each jet, the leading 200 jet constituents and their 4-momentum vector $(p_x,p_y,p_z,E)$ are included; jets with fewer than 200 constituents are zero-padded.

\subsection{Machine Learning Approaches}\label{sec:ml_jets}
Recently, many ML-based jet-taggers have been studied. Here, we highlight in particular a subset of these models that have demonstrated state-of-the-art (or near SotA) performance on the task of top tagging, and which represent jet data in a computationally efficient manner. There is a large body of work representing jets as images by mapping energy deposits in the calorimeter to image pixels (e.g.~\cite{jet_images}); this enables the use of common convolutional neural network (CNN) architectures like ResNeXt for jet tagging~\cite{resnext}. Similarly, there are multiple approaches representing jets as point clouds or graphs, enabling the use of GNNs for jet tagging; ParticleNet~\cite{particlenet} is an accurate and well-established GNN-based tagger that incorporates dynamic $K$-nearest-neighbor (KNN) graph building at each graph convolution stage. Energy Flow Polynomials (EFPs)~\cite{efps} are a set of jet substructure variables developed to form a linear basis for all infrared- and collinear-safe jet observables that can be efficiently computed using graph-theoretic representations; these EFPs can then be used as inputs to linear models for jet tagging. 

\subsection{Equivariant Graph Neural Networks}
Many real-world datasets, including data from HEP experiments, exhibit known symmetries. This motivates the development of ML architectures that can produce predictions that are invariant or equivariant with respect to symmetric transformations. Formally, for symmetry group $G$ with transformations $T_g:X\rightarrow X$ on the input space $X$, a model is defined as equivariant if it is composed of functions $\phi:X\rightarrow Y$ where there is some transformation $S_g$ on output space $Y$ such that $\phi(T_g(x))=S_g(\phi(x))$, for all $g\in G$. There is a large body of work exploring equivariance in ML models and we highlight those that are particularly relevant to our work. A general approach to group convolutions was first developed in \cite{groupconv}, specific methods to achieve E(3) or SE(3) equivariance via higher order group representations were presented in \cite{e3a,e3b,e3c,e3d}, and a simplified method using only scalar and vector quantities to enforce E(n) symmetry with respect to an input set of points was demonstrated in \cite{satorras2021en}, called EGNN. For the specific case of particle physics symmetries, a DNN with a layer exploiting Lorentz equivariant quantities was developed for jet tagging in \cite{lorentz_dnn} and a fully Lorentz-equivariant GNN relying on higher order tensor productes of the Lorentz group was presented in~\cite{lgn}. This Lorentz Group Equivariant Neural Network (LGN) uses finite-dimensional representations of the Lorentz group to create Lorentz-equivariant operators with learnable non-linearities that form a neural-network-based jet tagger that is fully equivariant to Lorentz group transformations\footnote{We also note a similar recent work LorentzNet~\cite{Gong:2022lye} released during preparation of this work. We are yet to compare VecNet with LorentzNet performance.}.

These equivariant architectures are of particular interest because of claims that they can provide simpler, generalizable, interpretable models. Specifically, in the case of HEP tasks like jet tagging, equivariant models may be substantially smaller and have reduced computational complexity and cost. However, it is typically difficult to directly quantify the impact of enforcing equivariance in a model because equivariant and non-equivariant models are generally developed separately and include other potentially confounding features.

\subsection{Tuning Equivariance with VecNet}

\begin{figure}[htb!]
    \centering
    \includegraphics[width=1\linewidth]{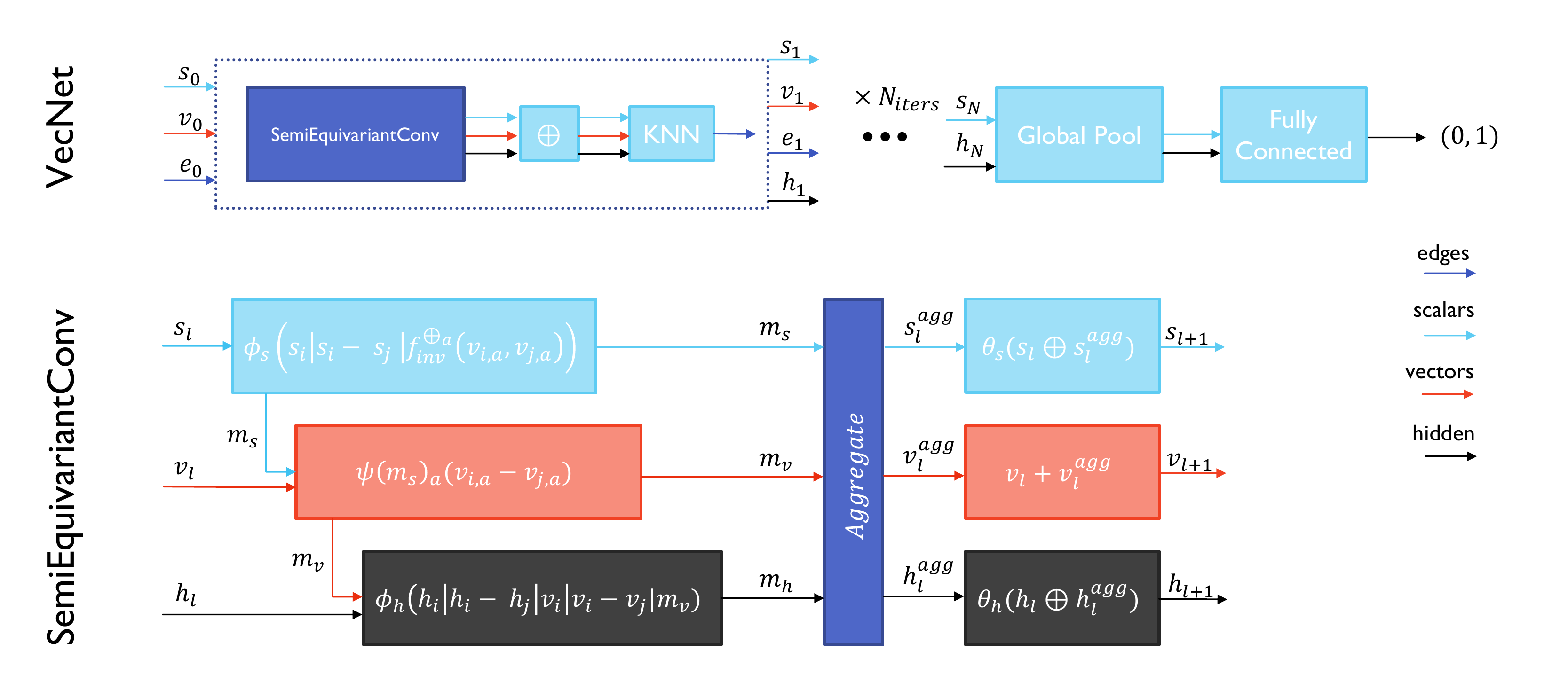}
    \caption{The generalized network VecNet, with configurable hyperparameters factored out. Equivariance is factored into the SemiEquivariant convolution, with scalar, vector and non-Lorentz channels included as an admixture, for message-passing iteration $l$.}
    \label{fig:general_net}
\end{figure}

In order to characterize the impact of equivariance and other GNN architecture choices, we present a generalized GNN for graph-level binary classification, called VecNet. The key contributions here are: a) A convolution for latent channels that can be tuned between a fully equivariant model, ``semi-equivariant", or fully non-equivariant; b) A graph construction hyperparameter that allows either fully connected graphs or static or dynamic graphs constructed in symmetry-invariant fashion; and c) A reusable implementation available publicly as part of the \textsc{Pytorch Geometric} library (the first such implementation of Euclidean/Lorentzian-equivariant GNN). The full model architecture is shown in \cref{fig:general_net}, which is generic to any desired symmetry group, but henceforth we focus on the Lorentz group $SO^+(1,3)$. Broadly, the model takes as input (Lorentz) scalar $s$ and vector $v$ quantities, and either a fully connected or KNN graph of edges $e$. KNN distances between vectors $v_0, v_1$ are given by the spacetime interval $|v_0^\mu - v_1^\mu|^2 = (v_0^0 - v_1^0)^2 - \sum_{i=1}^3(v_0^i - v_1^i)^2$. These values are passed through $N$ iterations of the SemiEquivariant convolution. They are aggregated at each step, a skip connection is (optionally) included, a new graph is (optionally) dynamically generated, and they are passed into the next iteration. A non-equivariant hidden feature channel $h$ is also passed through each convolution, and these are pooled at graph-level with the scalar channels to produce a final channel. This channel is passed through a fully-connected MLP to produce a class prediction for the jet. 

The SemiEquivariant convolution is based heavily on EGNN~\cite{satorras2021en}. For each node $i$ and its neighbors $j$, input scalar quantities $s_i, s_j$ are concatenated with any Lorentz-invariant quantities that can be formed between neighboring vector quantities $f_{inv}$. Here, we take this function to be the spacetime interval, given previously. These concatenated scalars are passed through an MLP $\phi_s$ and form the scalar messages $m_s$. To calculate vector messages, in the first iteration, each vector $v$ is cloned into vector channels $v_a$, which are subsequently independent. This gives the equivariance more expressiveness, and acts as a kind of attention mechanism in $a$. Vector messages are formed by weighting each vector message $(v_{i,a} - v_{j,a})$ by the scalar message attention head $\psi(m_s)_a$. $\psi$ is an MLP taking $m_s$ from an $N_s$-dimensional channel to $N_v$ channels - one for each attention head. Observe that $(v_{i,a} - v_{j,a})$ transforms as a Lorentz vector, so these channels remain equivariant throughout the network. Finally, hidden (non-equivariant) messages are formed by passing concatenated scalar and hidden features through an MLP $\phi_h$. Thus, if $N_v = N_s = 0$, this architecture limits to a non-equivariant ParticleNet-like model. If $N_h = 0$, the network limits to a fully-equivariant EGNN-like model.

\section{Hyperparameter Exploration}

To fairly estimate the sensitivity of jet tagging performance to innovations in models such as ParticleNet and LGN, we attempt to refactor all architectural decisions into configurable hyperparameters and conduct Bayesian hyperparameter optimization (HPO). We take two figures of merit (FOMs) as goals of the HPOs: the area under the efficiency-purity curve (the AUC) and the \textit{ant factor}. Given the \textit{background rejection rate} $\epsilon^{-1}_B$ as the inverse of the false positive rate $1/fpr$ for a particular signal efficiency (which we set to 30\%), we define the ant factor as
\begin{align*}
    ant &= \frac{\textnormal{accuracy}}{\textnormal{model size}} = \frac{\epsilon^{-1}_B}{N_{params}}
\end{align*}
The number of parameters $N_{params}$ in a model is the number of all weights and biases in all multi-layer perceptrons.

HPO is performed on the NERSC Perlmutter HPC cluster, which allows the parallel training of HP configurations on NVIDIA DGX A100 units, for approximately 40 GPU-days. We use a subset of training data for faster HP search: A [train, val, test] split of [600k, 40k, 40k] for large models (average channel width greater than 50), and a split of [200k, 40k, 40k] for narrower models. Narrower models benefit less from this extra data. The HP search space is reported in Appendix \cref{tab:hp_space}, and is explored with the W\&B Sweep API~\cite{wandb}.

\subsection{Hyperparameter Importance}
To understand which architectural choices impact performance most heavily, we calculate the HP sensitivity. Sensitivity, or \textit{importance}, is the by-product of training a random forest regression (RFR) algorithm, taking as input variables the HP configurations and the regressed output to be the FOM. We also report the positive or negative linear correlation between these FOMs and each hyperparameter. We perform the importance and linear fit coefficient calculations in \textsc{SciKit-Learn}~\cite{scikit-learn} after first normalizing hyperparameters and FOMs. These studies are summarised in \cref{fig:hp_importance}. From the table, we see that scaling-up hidden width and node layer depth generally improves AUC. That is, if there are no resource constraints, a large model consisting of non-equivariant channels of the sort of ParticleNet is likely to give SotA accuracy. Arbitrarily adding ``equivariance" (i.e. vector and scalar channels) generally penalises the model's accuracy. However, note that equivariant channels don't penalise the ant factor as much as adding hidden channels, as quantified by the linear correlation. This suggests that there is some midpoint of semi-equivariance for efficient models, which will be explored in the following sections.

\begin{table}
    \caption{Sensitivity results of hyperparameters to the AUC and ant factor. We report only those of significant importance ($>0.01$)}
    \label{fig:hp_importance}
    \centering
    \begin{tabular}{lcccc}
        \toprule
        Hyperparameter & \multicolumn{2}{c}{AUC} & \multicolumn{2}{c}{Ant Factor}\\
        \cmidrule(lr){2-3}\cmidrule(lr){4-5}
        & Importance & Coefficient & Importance & Coefficient \\ \midrule
Hidden width & 0.344 & 0.044 & 0.443 & -0.092 \\
Vector width & 0.193 & -0.055 & 0.232 & -0.049 \\
Scalar width & 0.194 & -0.005 & 0.226 & -0.056 \\
Batch norm & 0.017 & -0.006 & 0.016 & -0.01 \\
Layer norm & 0.029 & 0.006 & 0.034 & -0.001 \\
Num. node layers & 0.157 & 0.126 & 0.01 & 0.014 \\
Num. edge layers & 0.056 & 0.036 & 0.023 & 0.01 \\
LR decay factor & 0.012 & -0.057 & 0.016 & 0.086 \\
        \bottomrule
    \end{tabular}
\end{table}

\subsection{Model Performance}
Using the VecNet architecture, we achieve state-of-the-art performance\footnote{While the AUC achieved is slightly less than ParticleNet's SotA value, we remind that less than half of the full training dataset is used, to search a larger volume of HP space}, as measured by ROC AUC and background rejection rate. The HP configuration for this model is listed in Appendix \cref{tab:hp_space}. In order to characterize a high-performing model, we choose from the 433 HP configuration runs the 30 that are most similar to this highest-performing configuration. Within this bracket of configurations \{hidden width $> 80$, vector width $< 15$, scalar width $> 50$\}, performance is significantly higher than the mean AUC across the whole study. 

We first note that these 30 configurations vary widely - they are uniformly-distributed across graph construction type, node aggregation type, shortcut connection type, and number of graph iterations. Indeed, these show little importance in our sensitivity study, and it suggests that many of these architectural decisions are arbitrary and may add unnecessary complexity. In particular, dynamic KNN graph construction appeared to offer little benefit over a static construction at the start of training/inference. Of the 30 high-performing configurations, we also see some configurations with significantly higher background rejection rate than previous models, which we report in \cref{tab:tagger_comparison}. One is inclined to be skeptical of such high values. But we remind that the HPO is Bayes-optimizing towards an ant factor that depends on background rejection. This suggests that HPO is appropriate and useful for multiple objectives (i.e. model size and background rejection rate in this case).

\begin{table}
    \caption{Comparison of top quark jet taggers against the VecNet architecture. Due to VecNet's flexibility of equivariance, it offers a way to tune between accuracy and ant factor. We report two indicative configurations to highlight this.}
    \centering
    \begin{tabular}{llllll}
        \toprule
Model & Accuracy & AUC & $\epsilon_B^{-1}$ & $N_{params}$ & Ant \\ \midrule
ResNeXt & $0.936$ & $0.984$ & $1122\pm 47$ & $1.46$M & $0.0007$ \\
ParticleNet & $0.938$ & $0.985$ & $1298\pm46$ & $498$k & $0.0026$ \\
EFP & $0.932$ & $0.980$ & $384$ & $1$k & $0.384$ \\
LGN & $0.929$ & $0.964$ & $435\pm 95$ & $4.5$k & $0.097$ \\ \midrule
\multirow{2}{*}{VecNet (Ours)} & $0.935$ & $0.984$ & $4046$ & $633$k & $0.006$ \\ 
 & $0.931$ & $0.981$ & $3482$ & $15$k & $0.229$ \\ \bottomrule
    \end{tabular}
    \label{tab:tagger_comparison}
\end{table}

\subsection{Effect of Semi-Equivariance}

\begin{figure}
    \centering
    \begin{subfigure}[b]{0.48\textwidth}
        \includegraphics[width=\textwidth]{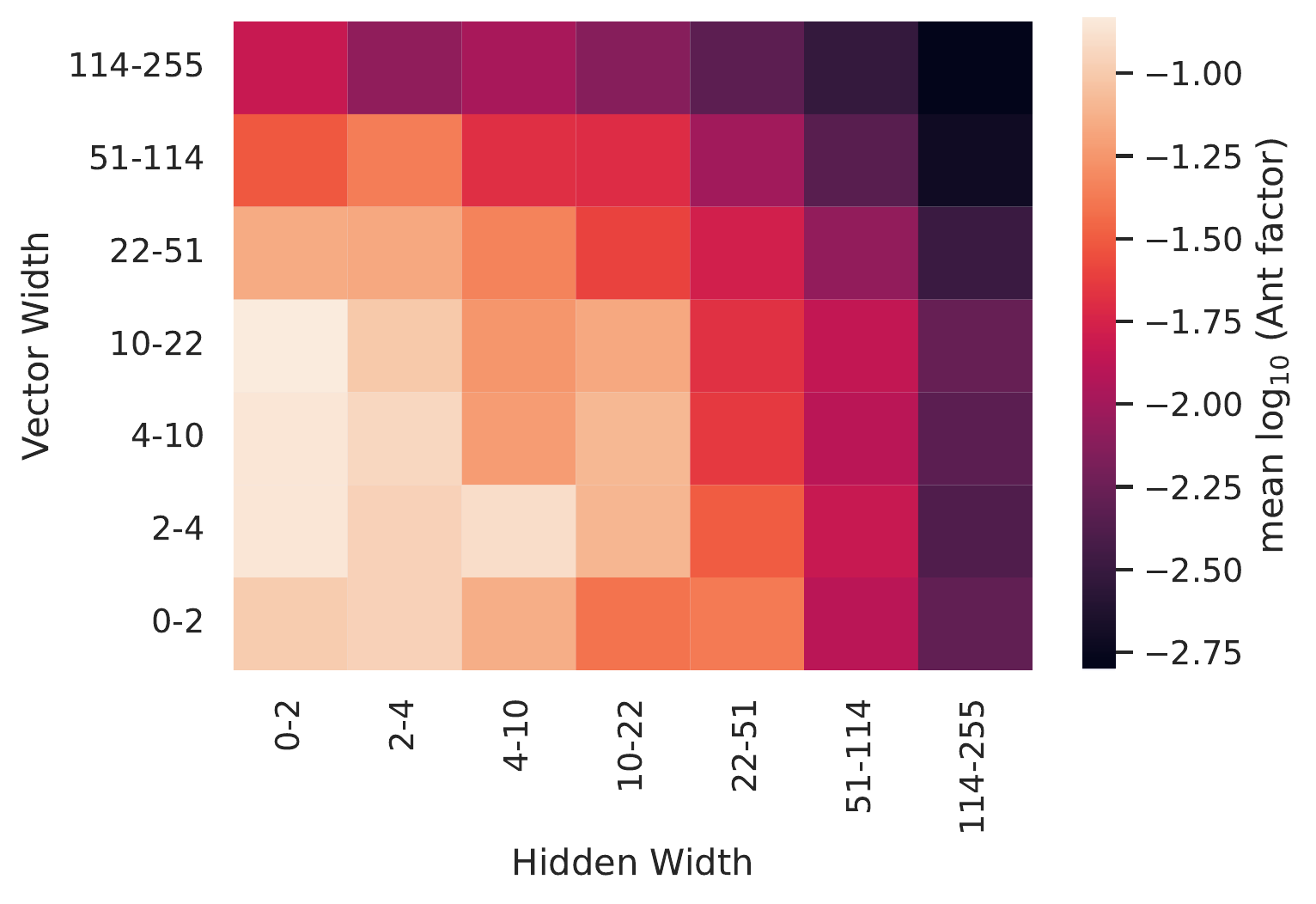}
        \caption{}
        \label{fig:channels_vs_ant}
    \end{subfigure}
    \hfill 
    \begin{subfigure}[b]{0.48\textwidth}
        \includegraphics[width=\textwidth]{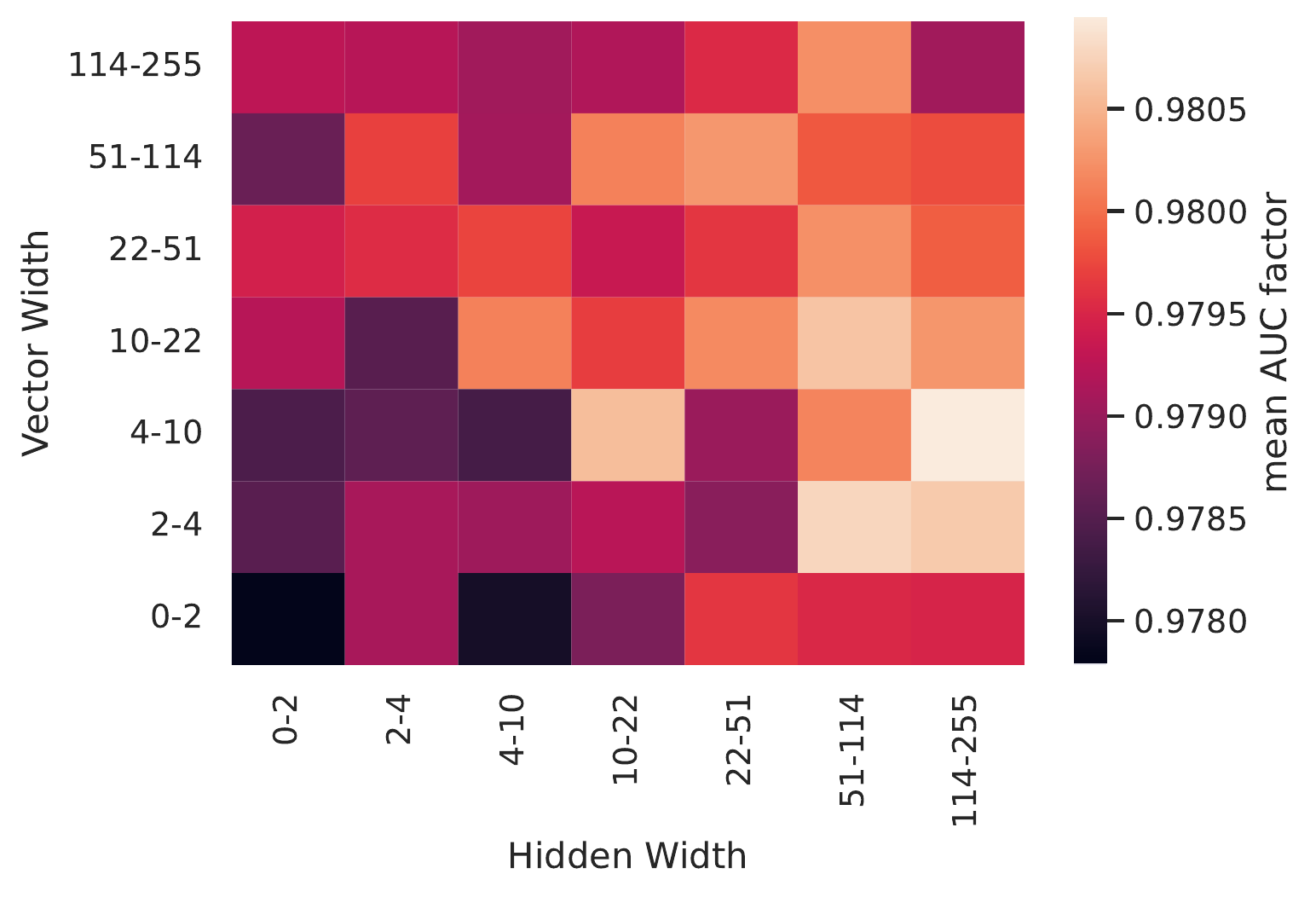}
        \caption{}
        \label{fig:channels_vs_auc}
    \end{subfigure}
    \caption{Impact of equivariance on the two FOMs considered in this paper: Ant factor and AUC. We search across high equivariance (top left of each grid) to low equivariance (bottom right of each grid). Binning is inclusive to the right, i.e. $(N_1, N_2]$}
\end{figure}

For a non-equivariant architecture, one would expect (and indeed we approximately observe this along the lowest bin row of \cref{fig:channels_vs_ant}) a monotonically increasing ant function - that is, a model's discriminating power increases as it grows in size, either faster or slower than the rate of parameter growth. Introducing equivariant channels makes this function more complex, and there appears to be a region of non-trivial optima between the edge cases of fully-equivariant and non-equivariant\footnote{Note that there is not simply ``more information" available in this area - even without vector channels, the hidden channels have access to all available features}. Indeed, the most resource-efficient configuration (given as the second example in \cref{tab:tagger_comparison}) was achieved with 4 hidden channels, 2 vector channels and 8 scalar channels.

This can be quantified using the sensitivity study. While generally the model accuracy and ant factor are highly sensitive to model size, when we focus on this region of optimal efficiency\footnote{Specifically hidden width $\in [3, 35]$, vector width $\in [3, 35]$} and re-fit an RFR, we see hidden and vector importances of $0.152$ and $0.095$ relative to ant factor. This lower sensitivity is consistent with some non-linear optimum. Interpretability studies remain to be done to explore how the hidden and vector channels are exploiting and sharing physically meaningful information.

\section{Conclusion and Future Work}

In this work, we have presented a generalized GNN architecture to allow interpolation between equivariant and non-equivariant learning. With different configurations, the same model can deliver SotA performance (AUC=0.984) with no equivariance, or almost-SotA performance (AUC=0.983) with 20 times fewer parameters by exploiting equivariant channels. We capture this trade-off with a novel figure of merit - the ant factor. Guided by this metric, we thus suggest that fully-equivariant models are neither the most accurate, nor the most resource-efficient. Instead, a semi-equivariant network can quickly learn symmetry-invariant physics with vector and scalar channels, while gaining last-mile expressiveness from hidden channels. In the future we plan to characterize the physical phenomena learned by this method of semi-equivariance, and particularly what information the GNN is able to learn beyond known Lorentzian and Euclidean symmetries. We are currently exploring GNN interpretability and explainable AI methods for this purpose.

\ack
This work is supported by IRIS-HEP through the U.S. National Science Foundation (NSF) under Cooperative Agreement OAC-1836650 and by Department of Energy grant DE-SC0007968. This research was supported in part by the Exascale Computing Project (17-SC-20-SC), a joint project of the Office of Science and National Nuclear Security Administration. This research used resources of the National Energy Research Scientific Computing Center (NERSC), a U.S. Department of Energy Office of Science User Facility located at Lawrence Berkeley National Laboratory, operated under Contract No. DE-AC02-05CH11231

\appendix
\section*{Appendix}

The full set of hyperparameters and their search space is detailed in Appendix \cref{tab:hp_space}. All are uniformly searched, with the exceptions of the scalar, vector and hidden channel widths, which are sampled from a log-uniform distribution. The highest-performing configuration as measured by AUC is given in parentheses and bold in the table.

\begin{table}[htb!]
\caption{Hyperparameter search space. Best configuration is boldened.}
     { \small
    \subfloat[Model hyperparameters and their ranges]{
    \begin{tabular}{ll}
        \toprule
 Hyperparameter & Values \\ \midrule
 $N$ edge layers $\phi_s, \phi_h$ & $[1, 3]$ \textbf{(3)}\\
$N$ node layers $\theta_s, \theta_h$ & $[0, 3]$ \textbf{(2)}\\
$N$ graph iterations & $[1, 6]$ \textbf{(3)}\\
$N$ channels $\phi_s, \psi, \phi_h$ & $[1, 256]$ $\textbf{(67,9,125)}$\\
Shortcut & None, \textbf{Skip}, Concat\\
Activations & \textbf{ReLU}, SiLU, Tanh\\
Batch norm & True, \textbf{False} \\ 
Layer norm & \textbf{True}, False \\ \bottomrule
    \end{tabular}} \quad \subfloat[Training hyperparameters and their ranges]{
    \begin{tabular}{ll}
        \toprule
Hyperparameter & Values \\ \midrule
\multirow{2}{*}{Graph construction} & Static KNN, Dynamic KNN \\
 & \textbf{Fully Connected} \\
$K$ neighbors & $[3, 32]$ \\
Dropout & $[0.01, 0.4]$ \textbf{(0.1)} \\
Learning rate & $[10^{-5}, 10^{-2}]$ \textbf{(0.003)} \\
$N$ epochs & $[10, 50]$ \textbf{(24)}\\ \bottomrule
    \end{tabular}}
    }
    \label{tab:hp_space}
\end{table}

\bibliographystyle{iopart-num}
\bibliography{refs}

\end{document}